\documentclass[a4paper]{article}
\usepackage{INTERSPEECH2022}
\usepackage[table,xcdraw]{xcolor}
\usepackage{multirow}
\title{Combination of Time-domain, Frequency-domain, and Cepstral-domain Acoustic Features for Speech Commands Classification}
\name{Yikang Wang$^1$, Hiromitsu Nishizaki$^1$}
\address{
  $^1$Integrated Graduate School of Medicine, Engineering, and Agricultural Sciences, \\ University of Yamanashi, Japan}
\email{wwm1995@alps-lab.org, hnishi@yamanashi.ac.jp}

\begin{document}

\maketitle
\begin{abstract}
  In speech-related classification tasks, frequency-domain acoustic features such as logarithmic Mel-filter bank coefficients (FBANK) and cepstral-domain acoustic features such as Mel-frequency cepstral coefficients (MFCC) are often used. However, time-domain features perform more effectively in some sound classification tasks which contain non-vocal or weakly speech-related sounds. We previously proposed a feature called bit sequence representation (BSR), which is a time-domain binary acoustic feature based on the raw waveform. Compared with MFCC, BSR performed better in environmental sound detection and showed comparable accuracy performance in limited-vocabulary speech recognition tasks. In this paper, we propose a novel improvement BSR feature called BSR-float16 to represent floating-point values more precisely. We experimentally demonstrated the complementarity among time-domain, frequency-domain, and cepstral-domain features using a dataset called Speech Commands proposed by Google. Therefore, we used a simple back-end score fusion method to improve the final classification accuracy. The fusion results also showed better noise robustness.
\end{abstract}

\section{Introduction}

In the past, the conventional automatic speech recognition (ASR) systems based on Hidden Markov Model (HMM) and Gaussian Mixture Models (GMM) \cite{zissman1993automatic,makowski2020voice}, usually used acoustic features designed from perceptual evidence for recognition and classification tasks, such as the Mel-frequency cepstral coefficients (MFCC) \cite{davis1980comparison,winursito2018improvement,lokesh2019speech}. As a cepstral-domain feature, MFCC can store the information of the corresponding spectral envelope to the low-frequency part of the pseudo-frequency axis, and the spectral envelope often contains the semantic component of speech. So using only 12-20 dimensional MFCC features not only expresses enough linguistic information but also helps to control the complexity of the HMM-GMM model. However, with the continuous development and wide application of deep neural network (DNN) technology in recent years, various neural network-based ASR systems, such as recurrent neural networks (RNN), convolutional neural networks (CNN), and graph neural networks (GNN) are widely used for speech-related tasks \cite{parcollet2019quaternion,gulati2020conformer,han2020contextnet,snyder2018x,baevski2020wav2vec,wu2020comprehensive,tzirakis2021multi}. Since there is no need to worry about higher feature dimensions leading to increasing model complexity, a frequency-domain feature, log Mel-filter bank coefficients (FBANK) \cite{deng2013recent,sarangi2020optimization,juvela2019gelp} is more frequently adopted. However, in other sound-related recognition and classification tasks other than speech, such as speaker discrimination \cite{qin2021simple,cai_exploring_2018}, environmental sound detection \cite{mushtaq2021spectral,guzhov2021esresnet}, and paralinguistic classification \cite{schuller2021interspeech,campbell2021paralinguistic}, the raw waveform is more favored by researchers because it retains all acoustic details. Several studies reported that it is better to use time-domain features such as raw waveform when using neural networks for audio classification because the length of the time series of the input data is longer, which allows deeper neural network models to be applied \cite{ravanelli2018speaker,bravo2021bioacoustic,purohit2018acoustic}.

Inspired by this, we proposed a time-domain acoustic feature, bit sequence representation (BSR) in our previous study \cite{okawa2019audio,9306262}, which preserves the information of the raw waveform completely and extends it to 2 dimensions, and demonstrated that the BSR has better performance and robustness than MFCC in environmental event recognition tasks \cite{okawa2019audio}. We also discussed the effects of maximum amplitude pre-processing in classification tasks \cite{9306262} for each feature in the GTZAN music/speech dataset \cite{tzanetakis2000marsyas} and the Speech Commands dataset proposed by Google \cite{warden2018speech}. 

In this paper, we conjecture that time-domain, frequency-domain, and cepstral-domain features are likely to be complementary in speech recognition or classification tasks. We design a simple features fusion experiment to evidence such complementarity. The classifiers' outputs are linearly combined to obtain the newly fused classification score, and these classifiers are trained separately from features in different domains. We also clearly show the specific differences in the classification results of two features using the confusion matrices. In addition, we propose a novel improvement BSR feature called Bit Sequence Representation of Half-Precision Binary Floating-Point Numbers (BSR-float16) to represent floating-point values more precisely.

We conduct experiments on the Speech Commands dataset \cite{warden2018speech} to confirm the differences between BSR-float16 and other features when it comes to speech classification tasks. The results showed that the proposed BSR-float16 feature in combination with other features improved the classification accuracy. The contributions of this paper are as follows:
\begin{itemize}
    \item We propose a novel improvement of BSR feature called BSR-float16 to represent the floating-point values more precisely.
	\item We design a simplest features fusion experiment, which uses a posterior probability-based scores combination method to evidence the complementarity among features in different domains. 
	\item We compare the classification accuracy of several acoustic features extracted under different signal-to-noise ratio (SNR) conditions and the accuracy after score fusion. The score improvement indicates that features in different domains are also complementary in noise robustness.
\end{itemize}
The remainder of this paper is organized as follows: in section 2, we present the transformation method of the BSR-float16 feature and the extraction of other features. In section 3, we introduce the back-end feature score linear combination method. In section 4, we discuss the experimental setup and results, and finally, we provide a conclusion of the paper and some directions for future work in Section 5.

\section{Features Extraction}
\subsection{BSR-float16} \label{sec2}
\begin{figure}[h!]
	\centering
	\includegraphics[width=0.6\linewidth]{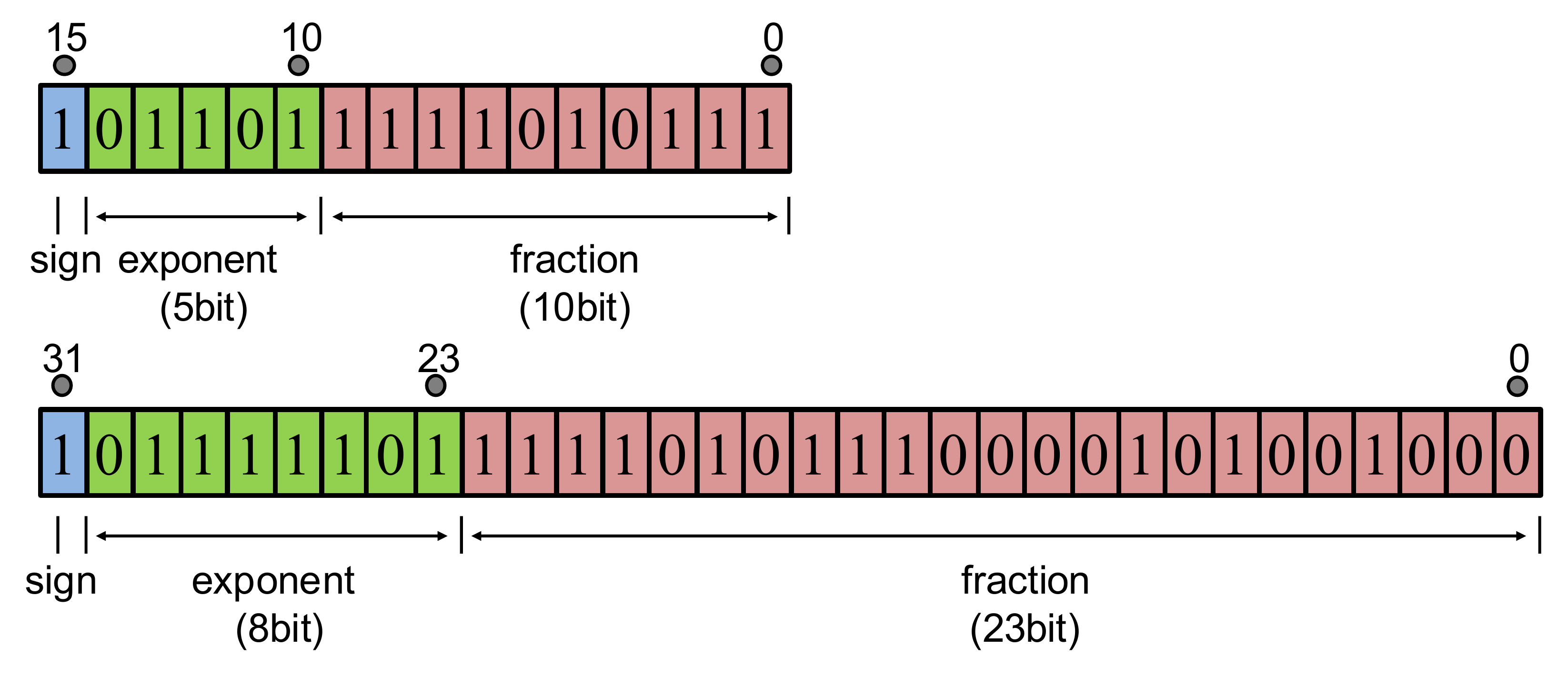}
	\caption{Floating-point number -0.49 representation on IEEE 754 half-precision binary floating-point format (float16)}
	\label{fig:IEEE754}
	\vspace{-0.3cm}
\end{figure}
In general, the sample value of an audio waveform is represented as a signed integer. For example, when the quantization bit rate is 16 bits per sample, the sampling value varies from -32768 to +32767. In our previous study \cite{okawa2019audio}, we investigated only the BSR of the integer values (BSR-int16). However, when we deal with raw waveforms, it is quite common for the dynamic range of amplitude to vary with each recorded waveform. In such cases, normalization is necessary to unify the ranges. When the raw waveforms are normalized, the sample values are mapped to floating-point values. Converting this to a BSR-int16 will lead to large rounding errors.
	
In this paper, we propose a novel BSR feature called BSR-float16, which transform normalized floating-point sample values without rounding error. This is achieved by following the IEEE 754 half-precision binary floating-point format. For example, assuming a sample point value is ``$-0.49$'', float16 for ``$-0.49$'' is displayed as ``[1, 0, 1, 1, 0, 1, 1, 1, 1, 1, 1, 1, 1, 1, 0, 1, 0, 1, 1, 1, 1].'' As shown in Figure~\ref{fig:IEEE754}, the bit representation is composed of three parts: the sign part, the exponent part, and the fraction part.

For IEEE 754 format \cite{8766229,goldberg1991every}, a floating-point number is transformed into a binary sequence by the following equation:
\begin{equation}\label{float16}
	\begin{aligned}[b]
		value=(-1)^{sign}\times2^{(E-15)}\times(1+\sum_{i=1}^{10}b_{10-i}2^{-i})
	\end{aligned}
\end{equation}
where the superscript $sign$ denotes the value of the sign bit, $E$ denotes the value of the decimal representation of the exponential part and $b_j$ denotes the binary value of the $j$-th bit. 

\begin{figure}[t]
	\centering
	\includegraphics[width=\linewidth]{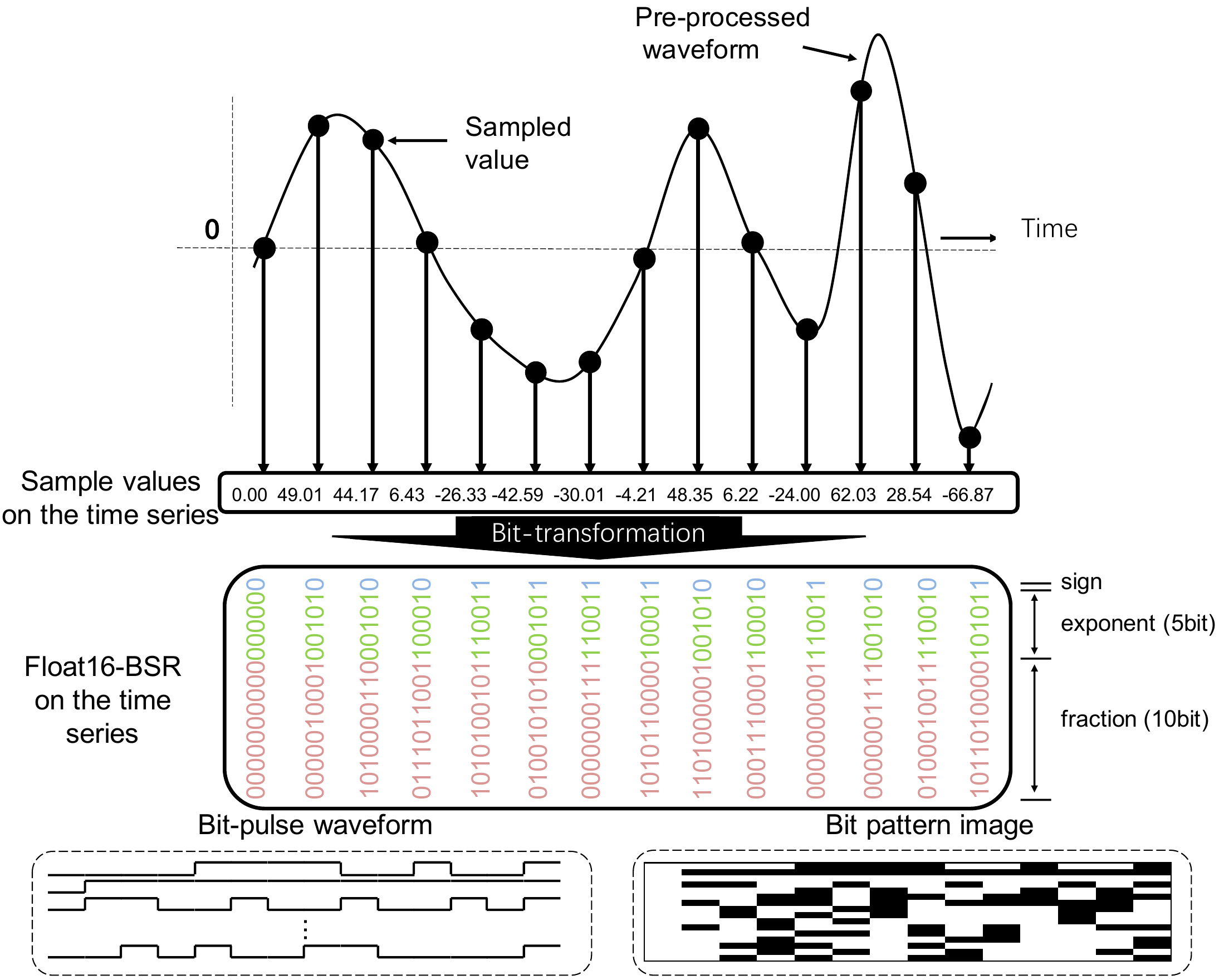}
	\caption{Transformation of a pre-processed floating-point audio waveform to BSR-float16. The blue, green, and red numbers are respectively corresponding to the sign, exponent, and fraction bits shown in Figure~\ref{fig:IEEE754}.}
	\label{fig:feature_extract}
	\vspace{-0.4cm}
\end{figure}

By expanding each sample point of an audio waveform into a 16-bit quantum bit vector, the original one-dimensional waveform has time-length T transformed to a two-dimensional feature shaped as $T \times 16$. Figure~\ref{fig:feature_extract} shows an example of the transformation from the pre-processed waveform to BSR-float16. After the BSR-based features are extracted, we have two ways to use the two-dimensional (2D) features, as follows:
\begin{enumerate}
	\item By extracting each dimension/digit of the bit representation vector sequence, we can obtain bit pulse. It enables such bit-pulse waveform input to the convolutional layers of a neural network-based classifier, where the one pulse is treated as one input channel. 
	\item By considering the whole 2D feature consisting of 0 and 1 as a picture, we can obtain a bit pattern image. We can process this bit pattern image like any other 2D feature, such as a photo image. 
\end{enumerate}
In this paper, we adopt the way in the lower left of Figure~\ref{fig:feature_extract}, which means inputting bit-pulse waveform to the classification model.
	
\subsection{Extraction of Other Features}

For comparison with BSR-float16, we chose FBANK and MFCC as representatives of their respective domains. In addition, we used the raw waveform-based classification results as our baseline.

\subsubsection{Raw Waveform}
\label{ssec:raww}
The raw waveform can be obtained directly by loading the WAVE file. Since the sampling rate of each file in the Speech Commands dataset is 16 kHz, 1s of PCM waveform is 16,000 samples. 

\subsubsection{FBANK}
\label{ssec:fbank}
The FBANK extraction is based on a pre-emphasized signal's short-term Fourier transform (STFT). A group of delta filters called Mel filter banks integrates the power spectrum. Multiply the STFT-transformed signal with the filters and take the logarithm to get the FBANK \cite{deng2013recent}. FBANK is a filtered frequency domain feature. Our experiments set the STFT window bin length to 0.025s and window step to 0.01s. 120-dimensional features were extracted, including 39 FBANK features, 1 energy feature, and their delta and double-delta features calculated from the log Mel spectrogram.

\subsubsection{MFCC}
\label{ssec:mfcc}
The MFCC features can be obtained by applying discrete cosine transform (DCT) from the FBANK.
As a cepstral-domain feature, MFCC contains information about the rate changes in the different spectrum bands. In other words, in the cepstral domain, the spectral envelope information is stored in the low-frequency part of the pseudo-frequency. In our experiments, 39-dimensional features, consisting of 12 statics MFCC features, 1 energy feature, and their delta and double-delta features are used. 
\vspace{-0.4cm}
\section{Back-end Score Fusion Method}

The feature combination method in this paper is based on linear score fusion of posterior probabilities. The models are first trained with features separately, and the test result scores of these models are linearly combined with others.
For example, if we train the classification models $f_a(~)$ and $f_b(~)$ base on feature $a$ and $b$, the prediction results of the two models for the input $X$ are $Y_a=argmax\bigl(f_a(X)\bigr)$ and $Y_b=argmax\bigl(f_b(X)\bigr)$ respectively, and the linear score fusion results $Y_{fus}$ become:
\begin{equation}\label{2cinbination}
Y_{fus}=argmax~\Bigl( \frac{1}{2}f_a(X) + \frac{1}{2}f_b(X)\Bigr)
\end{equation}
For the fusion score of three features:
\begin{equation}\label{3cinbination}
	Y_{fus}=argmax~\Bigl( \frac{1}{3}f_a(X)+\frac{1}{3}f_b(X)+\frac{1}{3}f_c(X)\Bigr)
\end{equation}

where $f(X)$ denotes the model $f(~)$ output probability distribution with input $X$. $Y$ denotes the final prediction of experiments.
Generally, we write the results of score fusion as:

\begin{equation}\label{Ncinbination}
	Y_{fus}=argmax~\Bigl( \sum_{i=1}^n\omega_if_i(X)\Bigr)
\end{equation}

where $n$ is the number of features, $f_i(~)$ denotes the corresponding model of the $i$-th feature, and $\omega_i$ represents the weight of the output probability score of $i$-th feature's model. Generally, $\sum_{i=1}^n\omega_i = 1$ and for this experiment $w_i = \frac{1}{n}$.

\begin{table*}[thb]
	\centering
	\caption{Classification accuracies [\%] of different combinations of features for various noise conditions in the test set. The bold numbers in the table indicate the best classification accuracy for different noise conditions of a single feature, two-feature combination, and three-feature combination, respectively, and the bold italic numbers indicate the best classification accuracy for each noise condition in all feature combinations.}
	\label{tbl:acc_expr2}
	\resizebox{1\textwidth}{!}{%
		\begin{tabular}{l|l|c|l|l|l|l|l|l|l|l|l}
			\toprule
			\multicolumn{2}{l|}{\multirow{2}{*}{Input Features}}                               & \multirow{2}{*}{Clear} & \multicolumn{3}{c|}{background noise} & \multicolumn{3}{c|}{white noise} & \multicolumn{3}{c}{pink noise} \\ 
			\multicolumn{2}{l|}{}                                                              &                           & 20dB        & 10dB       & 0dB        & 20dB      & 10dB      & 0dB      & 20dB      & 10dB     & 0dB      \\ \midrule
			\multirow{4}{*}{Single features}             & BSR-float16                          &            94.56               & 87.30       & 75.14      & 42.14      & 82.03     & 63.25     & 37.02    & 91.83     & \textbf{89.66}    & \textbf{68.43}    \\ 
			& MFCC                                 & 93.83                     & 87.66       & 77.50      & 41.59      & 87.48     & 69.47     & 26.86    & 91.65     & 86.21    & 64.42    \\
			& FBANK                                & \textbf{94.74}                     & \textbf{90.02}       & \textbf{80.40}      & \textbf{51.76}      & \textbf{88.75}     & \textbf{\textit{78.25}}     & \textbf{48.28}    & 92.74     & 89.11    & 65.15    \\
			& Raw waveform                         & 93.65                     & 85.30       & 75.86      & 33.09      & 77.13     & 45.89     & 12.16    & \textbf{94.19}     & 88.93    & 64.42    \\ \midrule
			\multirow{6}{*}{Two features Combinations}   & BSR-float16 \& MFCC                  & 95.83                     & 89.66       & 81.67      & 48.24      & 89.47     & 69.65     & 38.66    & 94.19     & 90.56    & \textbf{73.18}    \\
			& BSR-float16 \& FBANK                 & \textbf{96.55}                     & \textbf{90.93}       & 83.48      & \textbf{53.23}      & 88.93     & 75.87     & \textbf{\textit{50.27}}    & \textbf{\textit{95.28}}     & \textbf{\textit{92.92}}    & 72.81    \\
			& BSR-float16 \& Raw waveform        & 94.92                     & 88.38       & 80.04      & 42.14      & \textbf{\textit{92.40}}     & 56.31     & 18.15    & 94.01     & 90.20    & 69.16    \\
			& MFCC \& FBANK                        & 94.92                     & 90.74       & 82.76      & 53.23      & 91.11     & \textbf{77.51}     & 47.01    & 93.65     & 89.29    & 68.07    \\
			& MFCC \& Raw waveform              & 95.10                     & 88.75       & 80.22      & 43.81      & 87.11     & 61.06     & 13.79    & 93.83     & 91.11    & 68.43    \\ 
			& FBANK \& Raw waveform           & 95.28                     & 89.84       & \textbf{84.03}      & 51.02      & 88.02     & 70.93     & 26.50    & 95.10     & 92.20    & 71.53    \\ \midrule
			\multirow{4}{*}{Three features Combinations} & BSR-float16 \& FBANK \& Raw waveform &     \textbf{\textit{96.91}}                & 90.20       & 83.67      & 51.57      & 88.93     & 69.47     & 29.58    & 95.10     & \textbf{92.92}    & \textbf{\textit{74.45}}    \\ 
			& BSR-float16 \& FBANK \& MFCC         &   96.55                    & \textbf{\textit{91.65}}       & \textbf{\textit{84.94}}      & \textbf{\textit{54.90}}      & \textbf{91.29}     & \textbf{76.60}     & \textbf{49.36}    & \textbf{95.10}     & 92.01    & 73.54    \\
			& BSR-float16 \& MFCC \& Raw waveform  & 96.73                     & 89.84       & 82.94      & 47.50      & 88.38     & 63.62     & 19.06    & 93.47     & 91.47    & 72.99    \\ 
			& Raw waveform \& FBANK \& MFCC        & 96.01                     & 91.11       & 84.21      & 52.31      & 91.29     & 73.13     & 28.13    & 94.74     & 91.83    & 72.63    \\ \bottomrule
		\end{tabular}
	}
	\vspace{-0.4cm}
\end{table*}

\section{Experiment}
	\label{sec:expe}
\subsection{Dataset}
\label{ssec:dataset}
The Speech Commands dataset was proposed by Google in 2018. It consists of 35 different types of English command words, totaling 105,829 audio files. Each audio sample data encoded as linear 16-bit single-channel PCM values with 1s duration. In our experiment, 84,843 sound files are used as a training set, 9,981 sound files are used as a development set, and the remaining 11,005 files are used as a test set. This dataset also contains six one-minute duration noise files, which are encoded in the same way as the command word audios and can be divided into three categories: one is real environmental background noise, consisting of dishwashing, cat meowing, bicycle riding, and running tap noise. The remaining two categories are white noise and pink noise. 

To analyze the robustness of different features and their combinations, we combined the environmental sounds with all 105,829 clear command words audio files at 20 dB, 10 dB, and 0 dB SNR to synthesize noisy data. We also synthesized noisy data in the same way for the white and pink noise. Finally, We got nine synthesized noisy datasets.

\subsection{Model architectures}
\label{ssec:model}
\begin{figure}[tbh]
	\centering
	\includegraphics[width=0.65\linewidth]{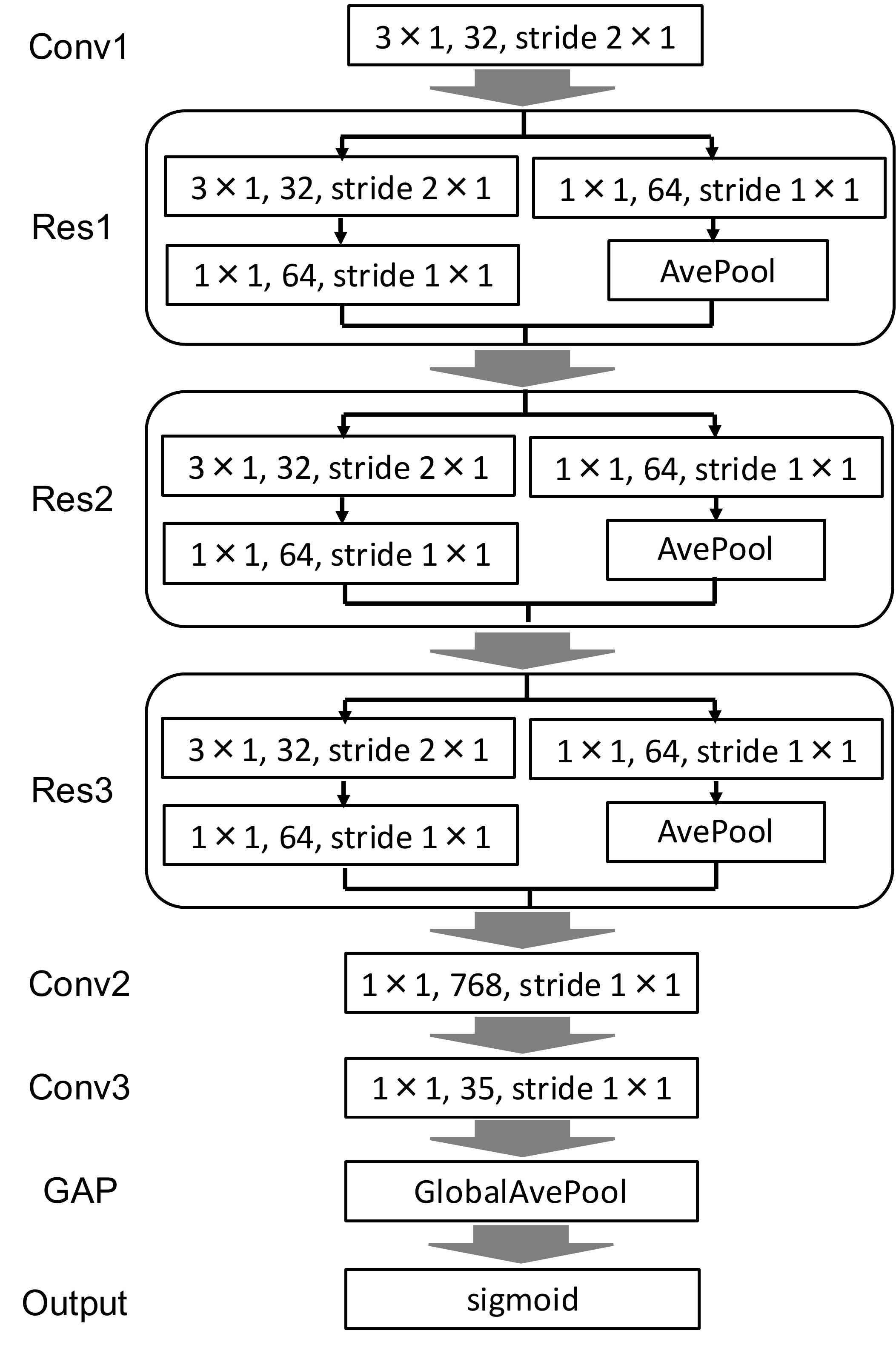}
	\caption{General construction of the ResNet model has 9 Conv. layers (Each Resnet block is considered to contain two convolutional layers.)}
	\label{fig:model}
	\vspace{-0.5cm}
\end{figure}

We referenced the structure of a novel ResNet model called Trident ResNet used in \textit{Task1} of the DCASE2020 \cite{suh2020designing}. We then tried to adjust the model depth and parameters so that each feature achieves the best classification performance, ensuring fairness in comparison.
As a general construction of the model, Figure~\ref{fig:model} is an example of a model with 9 Conv. layers.

\subsection{Experimental setup}
\label{ssec:setup}
First, to perform the speech command classification experiments, we padded all the command words files with zeros to equal lengths, so we could get time-domain features of length 16000 and frequency- and cepstral-domain features of length 99. We need to find the optimal model depth with the highest classification accuracy for each feature to ensure fairness of comparing. We varied the depth of the model by changing the number of ResNet blocks and placing each feature into all the models allowed by its respective length for training and validation. Eventually, we determined a model depth of 25 convolutional layers for baseline and BSR-float16, 9 convolutional layers for FBANK, and 11 convolutional layers for MFCC.

After that, the optimal model of each feature is trained on the train data set. Then we conducted noise robustness experiments by directly testing each model on the nine different noise condition datasets we mentioned in Section 4.1.

Finally, we used the method mentioned in section 3 to fuse the classification results of different features corresponding to different noise conditions. The final classification accuracy is placed in Table~\ref{tbl:acc_expr2}. 

We trained each model for 200 epochs with a mini-batch size of 16 to use GPU efficiently. Then we chose the ``momentumSGD'' as optimizer \cite{sutskever2013importance} with momentum 0.9 and learning rate schedule approach called ``SGDR,'' \cite{gotmare2018closer,loshchilov2016sgdr} which will restart with initial learning rate $lr$ at 5, 15, 35, 75, and 155 epochs. The value of $lr$ is 0.05 at 0 epoch, decreasing by 24\% for each restart.
\begin{figure}[thb!]
	\centering
	\includegraphics[width=0.92\columnwidth]{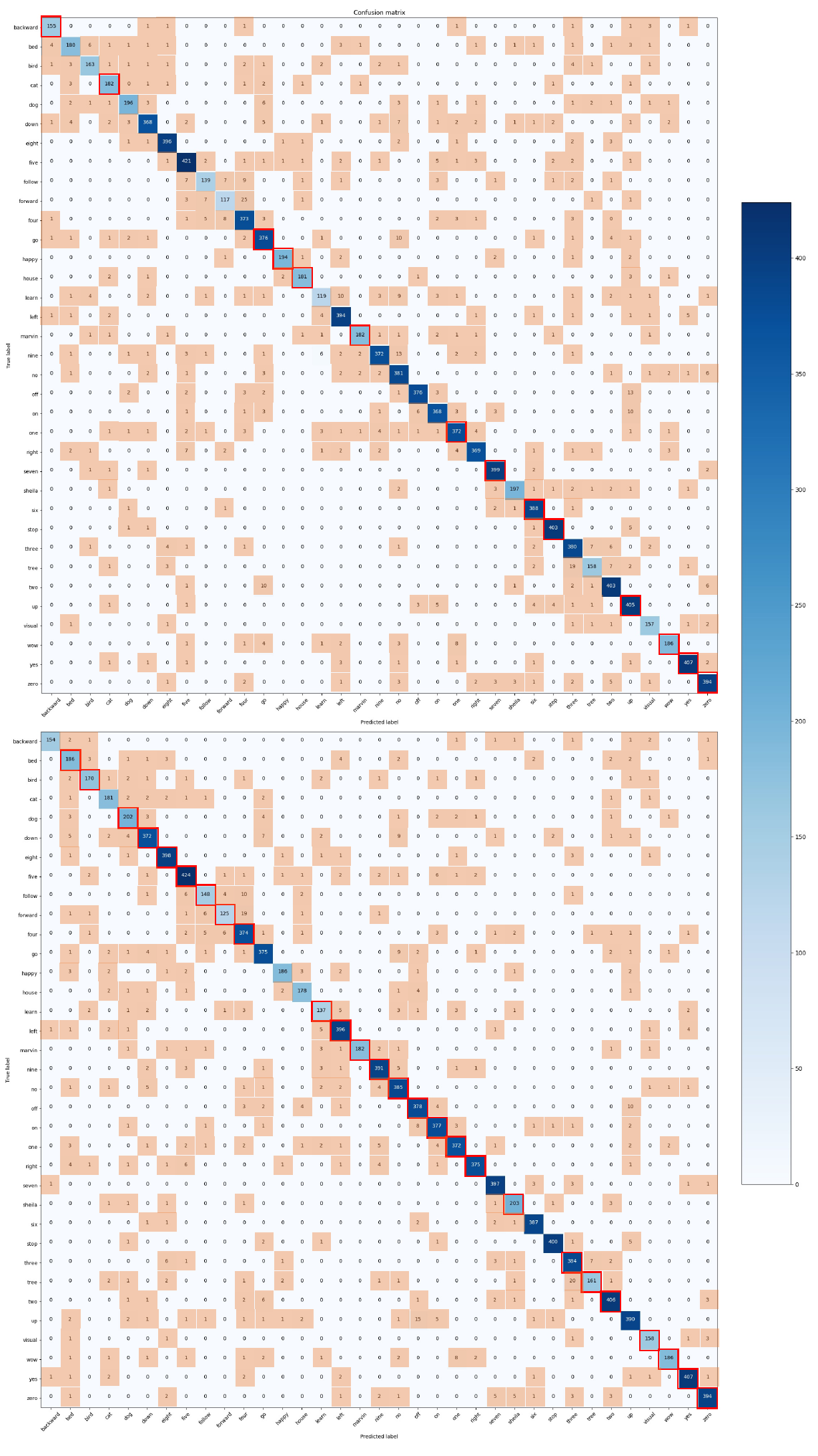}
	\caption{Comparison of confusion matrices based on the classification results of BSR-float16 (Top) and FBANK (Bottom)}
	\label{fig:cm}
	\vspace{-0.65cm}
\end{figure}

\subsection{Results and discussion}

Table~\ref{tbl:acc_expr2} shows the classification accuracy of different input features and their combinations under different noise conditions. Based on the results obtained from the single feature test under clear audio files, we plotted the corresponding confusion matrices. Figure~\ref{fig:cm} shows the confusion matrix of BSR-float16 and FBANK as an example. The horizontal axis of each confusion matrix indicates the classification labels, and the vertical axis indicates the predicted results. The red boxes indicate better results than the other acoustic feature. If a red box appears in the diagonal of the confusion matrix above, it means BSR-float16 has higher classification accuracy than FBANK for the corresponding word, and vice versa. The orange blocks indicate the location where the specific classification error of each feature occurs. By comparing the two matrices, it can be seen that BSR-float16 and FBANK have different tendencies in recognizing different words. This point supports the conclusion of our following fusion experiments.

It can be observed in Table 1 that the combination of BSR-float16, FBANK, and MFCC can significantly improve the final classification accuracy under the environmental background noise scenes. This result proves that the time-domain, frequency-domain, and cepstral-domain acoustic features are highly complementary in the clean-sound scenarios or the environmental background noise scenarios.

Moreover, the classification results about results with both white and pink noise in Table 1 show that good performance can be obtained by fusing the classification scores of FBANK and BSR-float16 only, indicating that MFCC is less complementary to the remaining two in the condition of synthetic noise. It is sure that in most cases, the noise robustness is improved after the score fusion. However, the possibility that the classification accuracy decreases after the fusion of scores cannot be excluded, for example, under 10 dB white noise conditions.

\section{Conclusions}

\label{sec:conclu}
This study aims to propose an improvement based on BSR, the BSR-float16 features, which enables a more accurate representation of floating-point numbers, and we demonstrate that BSR-float16 has comparable performance to other acoustic features such as MFCC and FBANK through a speech command classification task. Meanwhile, we conducted a simple linear feature score fusion method in the study, and the test results after score fusion under different noise conditions show that the classification accuracy was improved by 2.17 pt to 4.54 pt compared to a single feature. It is demonstrated that time-domain, frequency-domain, and cepstral-domain features are complementary in clean sound scenarios and environmental background noise scenarios. Since the BSR-float16 features can completely express the raw waveform information, we plan to use them in future research for tasks that require more paralinguistic information, such as speaker recognition. In addition, we hope to further investigate whether the complementarity of acoustic features from different domains can be generalized if we change tasks.

\section{Acknowledgments}
\label{sec:ack}

This work was supported by JSPS KAKENHI Grant Number 21H00901. 

\newpage
\bibliographystyle{IEEEtran}

\bibliography{mybib}

\begin{thebibliography}{10}
\providecommand{\url}[1]{#1}
\csname url@samestyle\endcsname
\providecommand{\newblock}{\relax}
\providecommand{\bibinfo}[2]{#2}
\providecommand{\BIBentrySTDinterwordspacing}{\spaceskip=0pt\relax}
\providecommand{\BIBentryALTinterwordstretchfactor}{4}
\providecommand{\BIBentryALTinterwordspacing}{\spaceskip=\fontdimen2\font plus
\BIBentryALTinterwordstretchfactor\fontdimen3\font minus
  \fontdimen4\font\relax}
\providecommand{\BIBforeignlanguage}[2]{{%
\expandafter\ifx\csname l@#1\endcsname\relax
\typeout{** WARNING: IEEEtran.bst: No hyphenation pattern has been}%
\typeout{** loaded for the language `#1'. Using the pattern for}%
\typeout{** the default language instead.}%
\else
\language=\csname l@#1\endcsname
\fi
#2}}
\providecommand{\BIBdecl}{\relax}
\BIBdecl

\bibitem{zissman1993automatic}
M.~A. Zissman, ``Automatic language identification using gaussian mixture and
  hidden markov models,'' in \emph{Proc. of International Conference on
  Acoustics, Speech, and Signal Processing (ICASSP)}, vol.~2.\hskip 1em plus
  0.5em minus 0.4em\relax IEEE, 1993, pp. 399--402.

\bibitem{makowski2020voice}
R.~Makowski and R.~Hossa, ``Voice activity detection with quasi-quadrature
  filters and gmm decomposition for speech and noise,'' \emph{Applied
  Acoustics}, vol. 166, p. 107344, 2020.

\bibitem{davis1980comparison}
S.~Davis and P.~Mermelstein, ``Comparison of parametric representations for
  monosyllabic word recognition in continuously spoken sentences,'' \emph{IEEE
  Transactions on Acoustics, Speech, and Signal Processing}, vol.~28, no.~4,
  pp. 357--366, 1980.

\bibitem{winursito2018improvement}
A.~Winursito, R.~Hidayat, and A.~Bejo, ``Improvement of mfcc feature extraction
  accuracy using pca in indonesian speech recognition,'' in \emph{Proc. of
  International Conference on Information and Communications Technology
  (ICOIACT)}.\hskip 1em plus 0.5em minus 0.4em\relax IEEE, 2018, pp. 379--383.

\bibitem{lokesh2019speech}
S.~Lokesh and M.~R. Devi, ``Speech recognition system using enhanced mel
  frequency cepstral coefficient with windowing and framing method,''
  \emph{Cluster Computing}, vol.~22, no.~5, pp. 11\,669--11\,679, 2019.

\bibitem{parcollet2019quaternion}
T.~Parcollet, M.~Ravanelli, M.~Morchid, G.~Linar{\`e}s, C.~Trabelsi,
  R.~de~Mori, and Y.~Bengio, ``Quaternion recurrent neural networks,'' in
  \emph{Proc. of International Conference on Learning Representations (ICLR)},
  2019.

\bibitem{gulati2020conformer}
A.~Gulati, J.~Qin, C.-C. Chiu, N.~Parmar, Y.~Zhang, J.~Yu, W.~Han, S.~Wang,
  Z.~Zhang, Y.~Wu, and R.~Pang, ``{Conformer: Convolution-augmented Transformer
  for Speech Recognition},'' in \emph{Proc. of Conference of the International
  Speech Communication Association (INTERSPEECH)}, 2020, pp. 5036--5040.

\bibitem{han2020contextnet}
W.~Han, Z.~Zhang, Y.~Zhang, J.~Yu, C.-C. Chiu, J.~Qin, A.~Gulati, R.~Pang, and
  Y.~Wu, ``{ContextNet: Improving Convolutional Neural Networks for Automatic
  Speech Recognition with Global Context},'' in \emph{Proc. of Conference of
  the International Speech Communication Association (INTERSPEECH)}, 2020, pp.
  3610--3614.

\bibitem{snyder2018x}
D.~Snyder, D.~Garcia-Romero, G.~Sell, D.~Povey, and S.~Khudanpur, ``X-vectors:
  Robust dnn embeddings for speaker recognition,'' in \emph{Proc. of
  International Conference on Acoustics, Speech and Signal Processing
  (ICASSP)}.\hskip 1em plus 0.5em minus 0.4em\relax IEEE, 2018, pp. 5329--5333.

\bibitem{baevski2020wav2vec}
A.~Baevski, Y.~Zhou, A.~Mohamed, and M.~Auli, ``wav2vec 2.0: A framework for
  self-supervised learning of speech representations,'' \emph{Advances in
  Neural Information Processing Systems}, vol.~33, pp. 12\,449--12\,460, 2020.

\bibitem{wu2020comprehensive}
Z.~Wu, S.~Pan, F.~Chen, G.~Long, C.~Zhang, and S.~Y. Philip, ``A comprehensive
  survey on graph neural networks,'' \emph{IEEE transactions on neural networks
  and learning systems}, vol.~32, no.~1, pp. 4--24, 2020.

\bibitem{tzirakis2021multi}
P.~Tzirakis, A.~Kumar, and J.~Donley, ``Multi-channel speech enhancement using
  graph neural networks,'' in \emph{Proc. of International Conference on
  Acoustics, Speech and Signal Processing (ICASSP)}.\hskip 1em plus 0.5em minus
  0.4em\relax IEEE, 2021, pp. 3415--3419.

\bibitem{deng2013recent}
L.~Deng, J.~Li, J.-T. Huang, K.~Yao, D.~Yu, F.~Seide, M.~Seltzer, G.~Zweig,
  X.~He, J.~Williams, Y.~Gong, and A.~Acero, ``Recent advances in deep learning
  for speech research at microsoft,'' in \emph{Proc. of International
  Conference on Acoustics, Speech and Signal Processing (ICASSP)}, 2013, pp.
  8604--8608.

\bibitem{sarangi2020optimization}
S.~Sarangi, M.~Sahidullah, and G.~Saha, ``Optimization of data-driven
  filterbank for automatic speaker verification,'' \emph{Digital Signal
  Processing}, vol. 104, p. 102795, 2020.

\bibitem{juvela2019gelp}
L.~Juvela, B.~Bollepalli, J.~Yamagishi, and P.~Alku, ``{GELP: GAN-Excited
  Linear Prediction for Speech Synthesis from Mel-Spectrogram},'' in
  \emph{Proc. of Conference of the International Speech Communication
  Association (INTERSPEECH)}, 2019, pp. 694--698.

\bibitem{qin2021simple}
X.~Qin, N.~Li, C.~Weng, D.~Su, and M.~Li, ``Simple attention module based
  speaker verification with iterative noisy label detection,'' in \emph{Proc.
  of International Conference on Acoustics, Speech and Signal Processing
  (ICASSP)}, 2022.

\bibitem{cai_exploring_2018}
W.~Cai, J.~Chen, and M.~Li, ``Exploring the {Encoding} {Layer} and {Loss}
  {Function} in {End}-to-{End} {Speaker} and {Language} {Recognition}
  {System},'' in \emph{Proc. of Speaker Odyssey}, 2018, pp. 74--81.

\bibitem{mushtaq2021spectral}
Z.~Mushtaq, S.-F. Su, and Q.-V. Tran, ``Spectral images based environmental
  sound classification using cnn with meaningful data augmentation,''
  \emph{Applied Acoustics}, vol. 172, p. 107581, 2021.

\bibitem{guzhov2021esresnet}
A.~Guzhov, F.~Raue, J.~Hees, and A.~Dengel, ``Esresnet: Environmental sound
  classification based on visual domain models,'' in \emph{Proc. of 25th
  International Conference on Pattern Recognition (ICPR)}.\hskip 1em plus 0.5em
  minus 0.4em\relax IEEE, 2021, pp. 4933--4940.

\bibitem{schuller2021interspeech}
B.~W. Schuller, A.~Batliner, C.~Bergler, C.~Mascolo, J.~Han, I.~Lefter,
  H.~Kaya, S.~Amiriparian, A.~Baird, L.~Stappen, S.~Ottl, M.~Gerczuk,
  P.~Tzirakis, C.~Brown, J.~Chauhan, A.~Grammenos, A.~Hasthanasombat,
  D.~Spathis, T.~Xia, P.~Cicuta, L.~J. Rothkrantz, J.~A. Zwerts, J.~Treep, and
  C.~S. Kaandorp, ``{The INTERSPEECH 2021 Computational Paralinguistics
  Challenge: COVID-19 Cough, COVID-19 Speech, Escalation \& Primates},'' in
  \emph{Proc. of Conference of the International Speech Communication
  Association (INTERSPEECH)}, 2021, pp. 431--435.

\bibitem{campbell2021paralinguistic}
E.~L. Campbell, R.~Y. Mes{\'\i}a, L.~Docio-Fernandez, and C.~Garc{\'\i}a-Mateo,
  ``Paralinguistic and linguistic fluency features for alzheimer's disease
  detection,'' \emph{Computer Speech \& Language}, vol.~68, p. 101198, 2021.

\bibitem{ravanelli2018speaker}
M.~Ravanelli and Y.~Bengio, ``Speaker recognition from raw waveform with
  sincnet,'' in \emph{Proc. of IEEE Spoken Language Technology Workshop (SLT)},
  2018, pp. 1021--1028.

\bibitem{bravo2021bioacoustic}
F.~J. Bravo~Sanchez, M.~R. Hossain, N.~B. English, and S.~T. Moore,
  ``Bioacoustic classification of avian calls from raw sound waveforms with an
  open-source deep learning architecture,'' \emph{Scientific Reports}, vol.~11,
  no.~1, pp. 1--12, 2021.

\bibitem{purohit2018acoustic}
T.~Purohit and A.~Agarwal, ``Acoustic scene classification using deep cnn on
  raw-waveform,'' \emph{Tech. Rep., DCASE2018 Challenge}, 2018.

\bibitem{okawa2019audio}
M.~Okawa, T.~Saito, N.~Sawada, and H.~Nishizaki, ``Audio classification of
  bit-representation waveform,'' in \emph{Proc. of Conference of the
  International Speech Communication Association (INTERSPEECH)}, 2019, pp.
  2553--2557.

\bibitem{9306262}
Y.~{Wang}, M.~{Okawa}, and H.~{Nishizaki}, ``Analysis of bit sequence
  representation for sound classification,'' in \emph{Proc. of Asia-Pacific
  Signal and Information Processing Association Annual Summit and Conference
  (APSIPA ASC)}, 2020, pp. 621--626.

\bibitem{tzanetakis2000marsyas}
G.~Tzanetakis and P.~Cook, ``Marsyas: A framework for audio analysis,''
  \emph{Organised sound}, vol.~4, no.~3, pp. 169--175, 2000.

\bibitem{warden2018speech}
P.~Warden, ``Speech commands: A dataset for limited-vocabulary speech
  recognition,'' \emph{arXiv preprint arXiv:1804.03209}, 2018.

\bibitem{8766229}
``{IEEE} standard for floating-point arithmetic,'' IEEE Std 754-2019 (Revision
  of IEEE 754-2008), pp. 1--84, 2019.

\bibitem{goldberg1991every}
D.~Goldberg, ``What every computer scientist should know about floating-point
  arithmetic,'' \emph{ACM computing surveys (CSUR)}, vol.~23, no.~1, pp. 5--48,
  1991.

\bibitem{suh2020designing}
S.~Suh, S.~Park, Y.~Jeong, and T.~Lee, ``Designing acoustic scene
  classification models with cnn variants,'' DCASE2020 Challenge, Tech. Rep.,
  2020.

\bibitem{sutskever2013importance}
I.~Sutskever, J.~Martens, G.~Dahl, and G.~Hinton, ``On the importance of
  initialization and momentum in deep learning,'' in \emph{Proc. of
  International conference on machine learning}, 2013, pp. 1139--1147.

\bibitem{gotmare2018closer}
A.~Gotmare, N.~S. Keskar, C.~Xiong, and R.~Socher, ``A closer look at deep
  learning heuristics: Learning rate restarts, warmup and distillation,'' in
  \emph{Proc. of International Conference on Learning Representations}, 2018.

\bibitem{loshchilov2016sgdr}
I.~Loshchilov and F.~Hutter, ``Sgdr: Stochastic gradient descent with warm
  restarts,'' \emph{arXiv preprint arXiv:1608.03983}, 2016.

\end{thebibliography}

\end{document}